\documentclass[final            
  ]
  {aipproc}

\layoutstyle{8x11double}
\usepackage{graphicx}
\usepackage{units}
\usepackage{amsmath}
\usepackage{amssymb}
\usepackage{xspace}

\def\EeV{\ifmmode {\mathrm{\ Ee\kern -0.1em V}}\else
                   \textrm{Ee\kern -0.1em V}\fi\xspace}%
\def\eV{\ifmmode {\mathrm{\ e\kern -0.1em V}}\else
                   \textrm{e\kern -0.1em V}\fi\xspace}%
\def\gcm{\ifmmode {\mathrm{g/cm}^2}\else
                   {g/cm$^2$}\fi\xspace}%
\def\Xmax{\ifmmode {X_\mathrm{max}}\else
                   {$X_\mathrm{max}$}\fi\xspace}%
\def\sigmaXmax{\ifmmode {\mathrm{RMS}(X_\mathrm{max})}\else
                   {RMS$(X_\mathrm{max})$}\fi\xspace}%
\def\meanXmax{\ifmmode {\langle X_\mathrm{max}\rangle}\else
                   {$\langle X_\mathrm{max}\rangle$}\fi\xspace}%
\newcommand{\depth}[1]{\unit[#1]{\gcm}}
\newcommand{\energy}[1]{\unit[$10^{#1}$]{\eV}}

\newcommand{\numberOfEvents}{3754\xspace}

\newcommand{\firstData}{December 2004\xspace}
\newcommand{\lastData}{March 2009\xspace}
\newcommand{\lowD}{\unit[(106$^{+35}_{-21}$)]{\gcm/decade}\xspace}
\newcommand{\highD}{\unit[(24$\pm$3)]{\gcm/decade\xspace}}
\newcommand{\breakE}{\energy{18.24\pm 0.05}\xspace}
\newcommand{\highRMS}{55}
\newcommand{\lowRMS}{26}

\begin{document}

\title{Analysis of Longitudinal Air Shower Profiles
       measured by the Pierre Auger Observatory}


\renewcommand\XFMauthorsandtwotext{ }

\author{M.\ Unger}{
        address={Karlsruher Institut f\"ur Technologie (KIT),
          Postfach 3640, D-76021 Karlsruhe, Germany}
}
\author{for the Pierre Auger Collaboration}{
        address={Observatorio Pierre Auger,
        Av. San Martin Norte 304, 5613 Malarg\"ue, Argentinia\\
full author list at \url{http://www.auger.org/archive/authors_2010_12.html}}
}

\begin{abstract}
  We describe the analysis of longitudinal air shower
  profiles as measured by the fluorescence detectors of
  the Pierre Auger Observatory and present the
  measurement of the depth of maximum of extensive air showers, $\Xmax$,
  with energies $\ge$ \energy{18}. The measured energy evolution of the
  average of $\Xmax$ and its fluctuations, \sigmaXmax,
  are compared to air shower simulations for different primary particles.
\end{abstract}

\maketitle


\section{Introduction}
The determination of the chemical composition of ultra-high energy
cosmic rays is essential to understand the origin of cosmic rays and
to interpret the features observed in the ultra-high energy cosmic ray
flux.  For instance, the observed hardening of the cosmic ray energy
spectrum at energies between \energy{18} and \energy{19}, known as the
'ankle', might either be a signature of the transition from galactic
to extragalactic cosmic rays or a distortion of a proton-dominated
extragalactic spectrum due to energy losses~\cite{ankle}.  Moreover,
the flux suppression observed
above~4$\cdot$\energy{19}~\cite{bib:gzkmeas} could be either due to
propagation effects~\cite{bib:gzk} (photopion production of primary
protons or photonuclear reactions of primary nuclei) or a signature of
the maximum injection energy of the sources~\cite{maxEnergy}.

There are several experimental methods to estimate the primary composition
from cosmic ray induced air showers. Within the Pierre Auger Observatory
(see~\cite{bib:auger} and \cite{bib:bruce})
the observation of the longitudinal shower development
with fluorescence detectors allows to measure the depth of the maximum
of the shower evolution, $\Xmax$, which is sensitive to the primary mass\footnote{
For other methods based on the surface detector see e.g.~\cite{bib:compoLodz}.}.

With the generalization of Heitler's model of electron-photon cascades
to hadron-induced showers~\cite{bib:heitlerModel} and the
superposition assumption for nuclear primaries of mass $A$, the
average depth of the shower maximum,
\meanXmax, at a given energy $E$ is expected to follow
\begin{equation}
   \meanXmax = \alpha\left(\ln E-\langle\ln A\rangle\right) + \beta,
\label{eq:meanXmax}
\end{equation}
where $\langle\ln A\rangle$ is the average of the logarithm of the primary
masses.  The coefficients $\alpha$ and $\beta$ depend on the nature of
hadronic interactions, most notably on the multiplicity, elasticity
and cross-section in ultra-high energy collisions of hadrons with air,
see e.g.\ \cite{Ulrich:2009hm}.  The change of \meanXmax per decade of
energy is called {\itshape elongation rate}~\cite{bib:elongationRate},
$D_{10}=\frac{\mathrm{d}\meanXmax}{\rm{d} \lg E}$,
and it is sensitive to changes in composition with energy.
A complementary composition-dependent observable
is the magnitude of the shower-to-shower fluctuations
of the depth of maximum, \sigmaXmax,
which is expected to decrease with the number of primary
nucleons $A$ (though not as fast as $1/\sqrt{A}$
\begin{figure}[!b]
\centering
\includegraphics[clip,bb=0 16 560 503,width=0.732\linewidth]{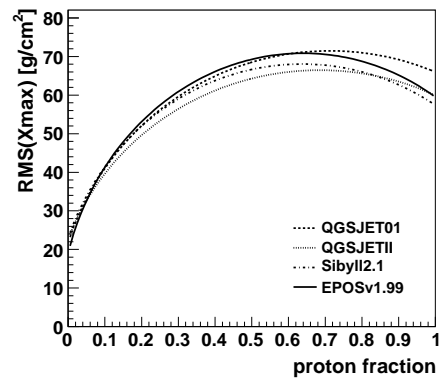}
\caption{\sigmaXmax  from different
         hadronic interaction models~\cite{bib:simulations} and a two-component
         p/Fe composition model ($E=$\energy{18}).}
\label{fig_RMS}
\end{figure}
\begin{figure*}[!t]
\centering
\includegraphics[clip,bb=0 -30 481 583,width=0.57\linewidth]{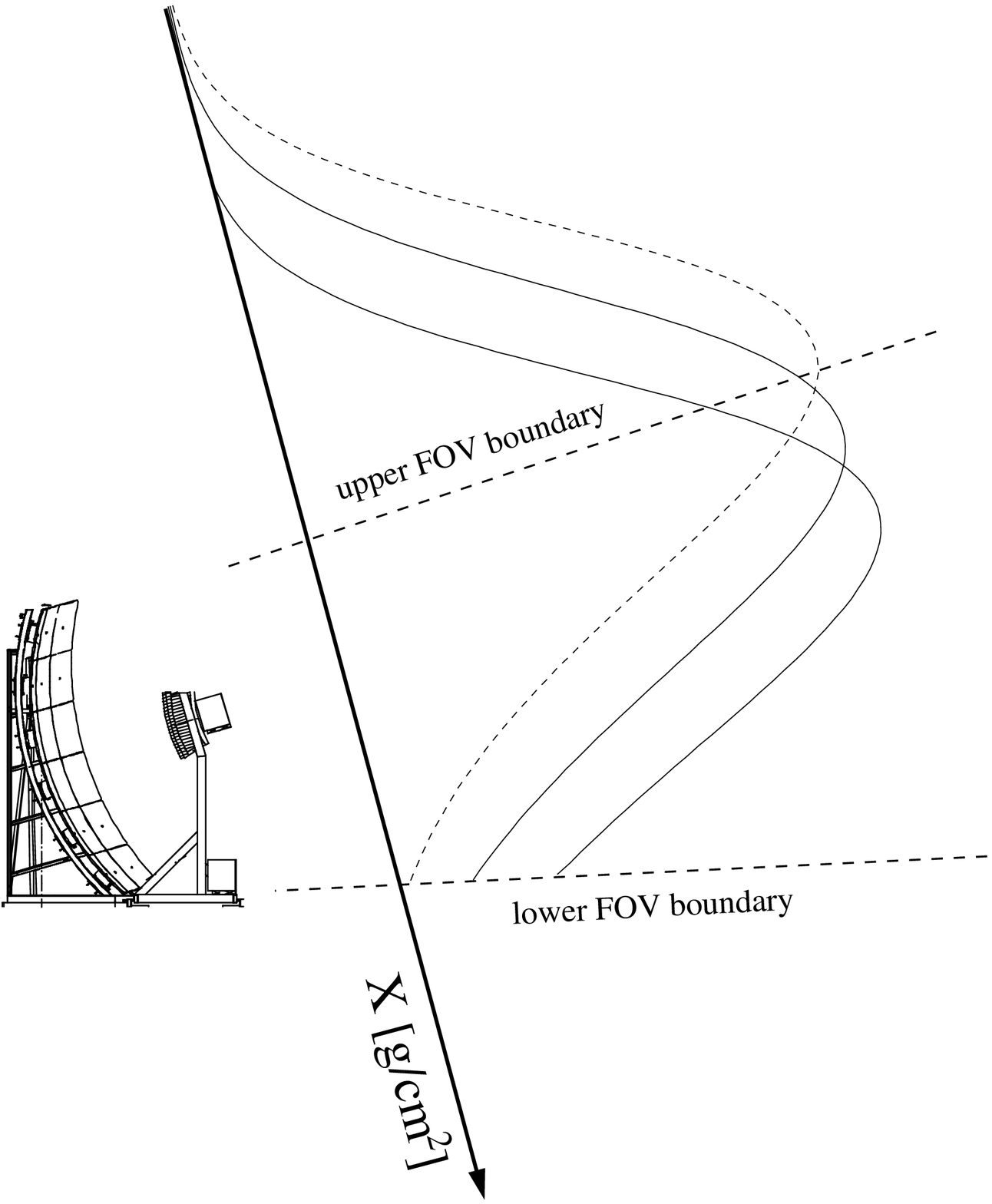}
\includegraphics[clip,bb=-39 -20 575 424,width=1.17\linewidth]{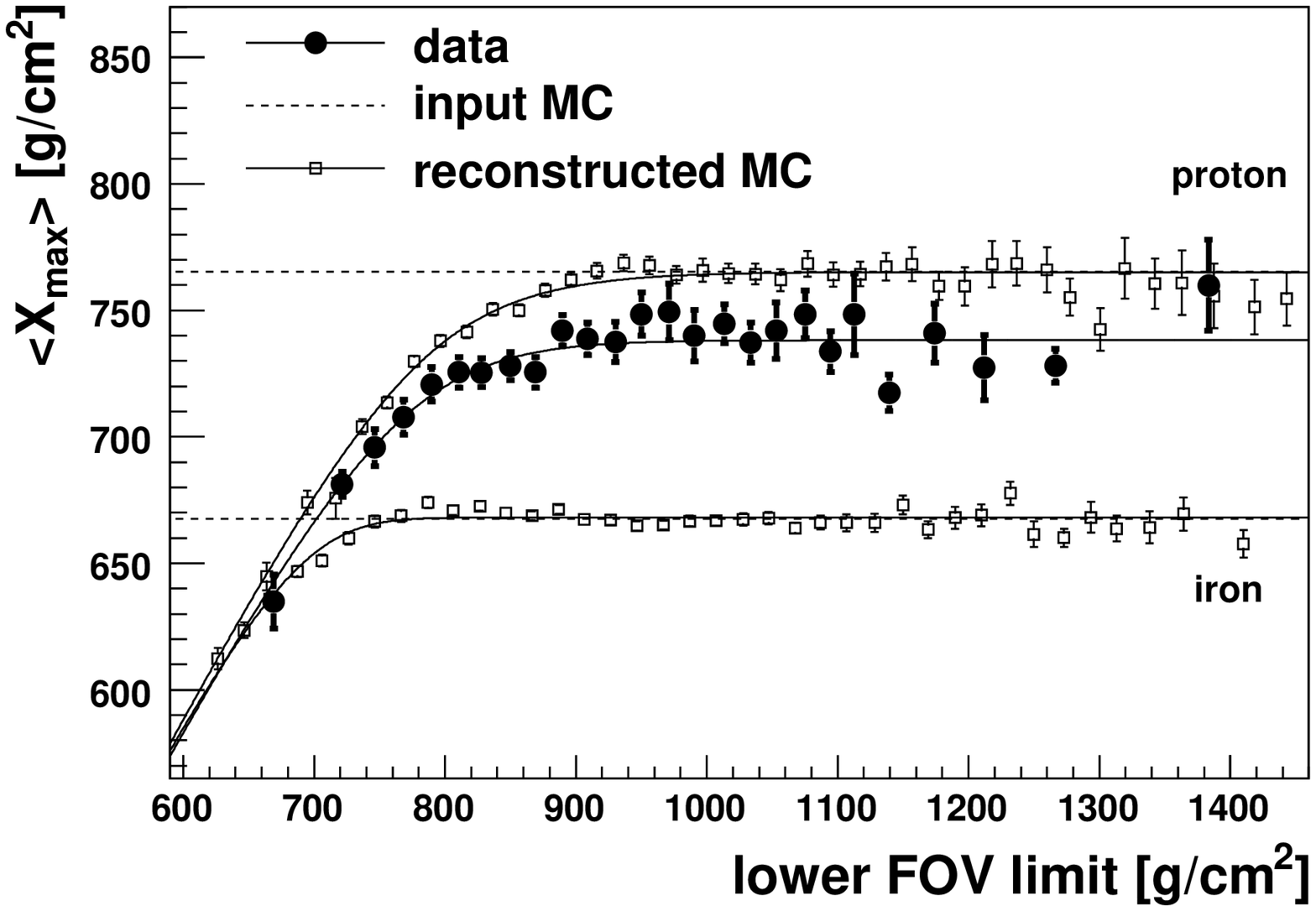}
\caption{Left: Illustration of the field of view bias. Right:
  Dependence of \meanXmax on the field of view boundary close
  to ground for data and MC (\energy{18.4}$<E<$\energy{18.6}).}
\label{fig_fov}
\end{figure*}
\cite{bib:fluctuations}) and to increase with the interaction
length of the primary particle. In case of a mixed composition, the full
width of the \Xmax distribution follows from the shower-to-shower fluctuations
of the individual mass groups and their separation in \meanXmax~\cite{bib:linsleyCompo}.
For a simple two-component composition of primaries
with masses $A_1$ and $A_2$ and abundances $f_1$ and $f_2=(1-f_1)$, \sigmaXmax is given by
\begin{equation}
   \sigmaXmax = \left(f_1 V_1 + f_2 V_2 + f_1 f_2 \Delta_X^2\right)^{\frac{1}{2}}
\label{eq:rms}
\end{equation}
where $V_i=\sigmaXmax_i^2$ and $\Delta_X=\meanXmax_1-\meanXmax_2$.
As can be seen in Fig.~\ref{fig_RMS}, a proton/iron mixture gives rise
to a broad maximum in \sigmaXmax for
$0.3\lesssim f_\mathrm{p}\lesssim 1$ and a rapid decrease of the width
towards $f_\mathrm{p}=0$.
\section{Data Analysis}
The $\Xmax$ results which were discussed at this workshop are based
on~\cite{bib:Abraham:2010yv} and use air shower data recorded between
\firstData and \lastData.  Only events detected in hybrid
mode~\cite{bib:hybrid} are considered, i.e.\ the shower development
must have been measured by the fluorescence detector (FD), and at
least one coincident surface detector station is required to provide a
ground-level time. Using the time constraint from the surface
detector, the shower geometry can be determined with an angular
uncertainty of 0.6$^\circ$~\cite{bib:angReso}.  The longitudinal
profile of the energy deposit is reconstructed~\cite{bib:profileRec}
from the light recorded by the FD using the fluorescence and Cherenkov
yields and lateral distributions from~\cite{bib:lightyields}.
With the
help of data from atmospheric monitoring devices~\cite{bib:augeratmo}
the light collected by the telescopes is corrected for the attenuation
between the shower and the detector and the longitudinal shower
profile is reconstructed as a function of atmospheric depth. \Xmax is
determined by fitting the reconstructed longitudinal profile with a
Gaisser-Hillas function~\cite{bib:gaisser-hillas}.

To assure a good $\Xmax$ resolution, the following quality cuts are
applied:
The impact of varying atmospheric conditions on the \Xmax measurement
is minimized by rejecting time periods with cloud
coverage and by requiring reliable measurements of the
vertical optical depth of aerosols. Profiles that are distorted by
residual cloud contamination are rejected by a loose cut on the
quality of the profile fit ($\chi^2$/Ndf$<$2.5).
We take into account events only with energies above \energy{18} where
the probability for at least one triggered surface detector station is 100\%,
irrespective of the mass of the primary particle~\cite{bib:hybridExposure}.
The geometrical reconstruction of showers with a large apparent
angular speed of the image in the telescope is susceptible to
uncertainties in the time synchronization between the fluorescence
and surface detector.
Therefore, events with a light emission angle towards the FD that is
smaller than 20$^\circ$ are rejected.  This cut also removes events
with a large fraction of Cherenkov light.
The energy and shower maximum can be reliably measured only if \Xmax
is in the field of view (FOV) of the telescopes (covering
$1.5^\circ$ to $30^\circ$ in elevation). Events for which only the
rising or falling edge of the profile is detected are not used.
Moreover, we calculate the expected statistical uncertainty of the
reconstruction of \Xmax for each event, based on the shower geometry
and atmospheric conditions, and require it to be better than \depth{40}.

\begin{figure*}[!t]
\centering
\includegraphics[width=0.89\textwidth]{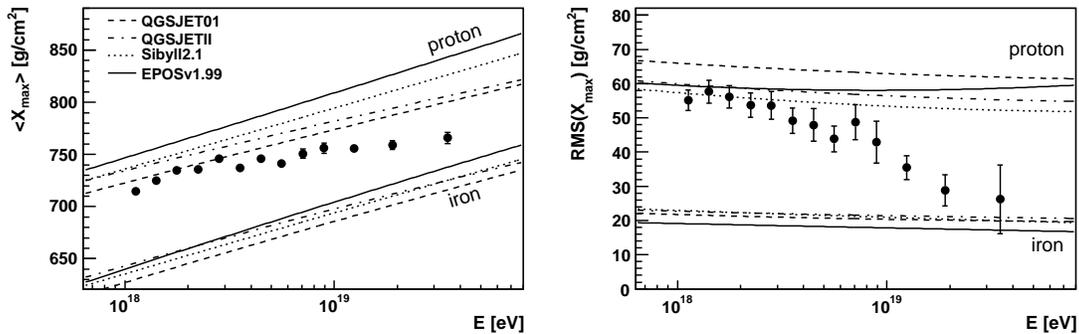}
\caption{\meanXmax and \sigmaXmax  compared
         with air shower simulations~\cite{bib:conex} using different
         hadronic interaction models~\cite{bib:simulations}.}
\label{fig_meanAndRMS}
\end{figure*}
The latter two selection criteria may cause a systematic under-sampling
of the tails of the \Xmax distribution, since showers developing very
deep or shallow in the atmosphere might be rejected from the data
sample (see illustration in the left panel of Fig.~\ref{fig_fov}).
To avoid a corresponding bias in the measured \meanXmax and
\sigmaXmax we apply fiducial volume cuts on the viewable \Xmax
range. For this purpose the effective upper and lower field of view
limits are calculated for each event and \meanXmax is measured as a
function of these limits. An example of the \meanXmax dependence on
the lower FOV limit is shown in Fig.~\ref{fig_fov} for data and
simulated events.  As can be seen, the \meanXmax is asymptotically
unbiased for events with a sufficiently deep FOV limit, but it is
systematically too shallow when the FOV boundary starts cutting into
the tails of the \Xmax distribution. Obviously, the unbiased region
depends on the distribution itself and can thus not be determined by
simulations.  Instead, we fit the data with the mean of a one-sided
truncated normal distribution (shown as solid lines in
Fig.~\ref{fig_fov}) and reject all events that have a FOV limit for
which the measured \meanXmax departs by more than \depth{5} from its
asymptotic value.

After all cuts, \numberOfEvents events are selected for the \Xmax
analysis. The \Xmax resolution as a function of energy for these
events is estimated using a detailed simulation of the FD and the
atmosphere. The resolution, defined by the full standard deviation, is
at the \depth{20} level above a few \EeV.  The difference between the
reconstructed \Xmax values in events that had a sufficiently high
energy to be detected independently by two or more FD stations is used
to cross-check these findings and as it was shown
in~\cite{bib:Abraham:2010yv}, the simulations reproduce the data well.

\section{Results}
The measured \meanXmax and \sigmaXmax
are measured in energy bins of
$\Delta\lg E=0.1$ below 10~\EeV and $\Delta\lg E=0.2$ above that energy.
The last bin starts at \energy{19.4}, integrating up to the highest
energy event ($E=(59\pm8)$~\EeV).  The systematic uncertainty of the
FD energy scale is 22\%~\cite{bib:icrcflux}.  Uncertainties
of the calibration, atmospheric conditions, reconstruction and event
selection give rise to a systematic uncertainty of $\le$\depth{13} for
$\meanXmax$ and $\le$\depth{6} for the RMS.
The results were found to be independent of zenith angle, time periods
and FD stations within the quoted uncertainties.

The measured values are displayed in Fig.~\ref{fig_meanAndRMS}.
A fit of $\meanXmax$
data with a constant elongation rate does not describe our data
($\chi^2$/Ndf=34.9/11), but
using two slopes yields a satisfactory fit ($\chi^2$/Ndf=9.7/9)
with an elongation rate of \lowD below \breakE and \highD above this
energy. If the properties of hadronic interactions do not change
significantly over less than two orders of magnitude in primary energy
($<$ factor 10 in center of mass energy), this change of $\Delta
D_{10}=$\unit[$(82^{+35}_{-21})$]{\gcm/decade} would imply a change in
the energy dependence of the composition around the ankle, supporting
the hypothesis of a transition from galactic to extragalactic cosmic
rays in this region.
\begin{figure*}[!t]
\centering
\includegraphics[width=0.89\textwidth]{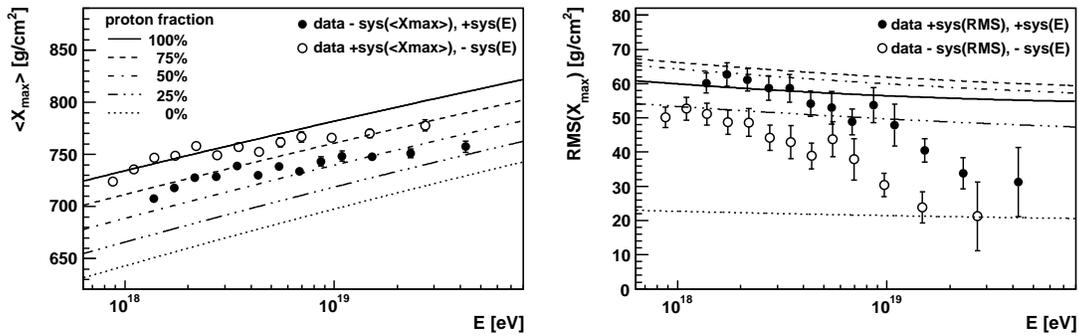}
\caption{\meanXmax and \sigmaXmax  compared
         to {\scshape QgsjetII} predictions for a two-component iron/proton
         composition. Data points show extreme values obtained by shifting
         the measurements by $\pm$ one sigma of the systematic uncertainties.}
\label{fig_Mix}
\end{figure*}

The shower-to-shower fluctuations, \sigmaXmax, are obtained by
subtracting the detector resolution in quadrature from the width of
the observed \Xmax distributions resulting in a correction of
$\le$\depth{6}. As can be seen in the right panel of
Fig.~\ref{fig_meanAndRMS}, we observe a decrease in the
fluctuations with energy from about \highRMS~to \depth{\lowRMS}
as the energy increases.  Assuming again that the hadronic interaction
properties do not change much within the observed energy range, these
decreasing fluctuations are an independent signature of an increasing
average mass of the primary particles.

For the interpretation of the absolute values of \meanXmax and
\sigmaXmax a comparison to air shower simulations is needed.  As can
be seen in Fig.~\ref{fig_meanAndRMS}, there are considerable
differences between the results of calculations using different
hadronic interaction models. These differences are not necessarily
exhaustive, since the hadronic interaction models do not cover the
full range of possible extrapolations of low energy accelerator
data. If, however, taken at face value, the comparison of the data and
simulations leads to the same conclusions as above, namely a gradual
increase of the average mass of cosmic rays with energy up to 59 \EeV.

It is illustrative to compare the data with predictions for
a simple two-component proton/iron model using the {\scshape QgsjetII} hadronic
interaction model. $\meanXmax$ and $\sigmaXmax$
simulations for different proton fractions $f$ are shown in Fig.\ref{fig_Mix}.
As can be seen, the $\meanXmax$ values change linearly with $f$ as expected
from Eq.~(\ref{eq:meanXmax}), whereas the width of the $\Xmax$ distribution
is very similar for proton fractions $f\gtrsim 0.3$.
To visualize the systematic uncertainties of the data, in this figure we
shifted the default results from Fig.~\ref{fig_meanAndRMS} by their systematic
uncertainties. Note that the
systematics on $\sigmaXmax$ are dominated by
the event selection, whereas the systematics on $\meanXmax$ are mainly
due to reconstruction uncertainties and atmospheric effects. Therefore
sys$(\meanXmax)$ and sys$(\mathrm{RMS})$ are uncorrelated and can be
 shifted independently.
Within this simplistic two-component model,
the data is compatible with a light or mixed composition
at low energies. At high energies, a heavy composition would result,
but the $\meanXmax$ would indicate a larger proton fraction than $\sigmaXmax$.
At high energies, this model corresponds to a heavy composition, however,
the $\meanXmax$ would indicate a larger proton fraction than $\sigmaXmax$.



\end{document}